\documentclass{PoS}

\usepackage{graphicx}
\usepackage{dcolumn}
\usepackage{bm}
\usepackage{amssymb}
\usepackage{xcolor}
\usepackage{amsmath}

\title{Dark Matter at the Intensity Frontier: the new MESA electron accelerator facility}

\ShortTitle{Dark Matter at the Intensity Frontier}

\author{\speaker{Luca Doria}
        \thanks{Full affiliation: PRISMA Cluster of Excellence and Institut f\"ur Kernphysik,
          Johannes Gutenberg-Universit\"at Mainz, Johann-Joachim-Becher-Weg 45 D 55128 Mainz}\\
        PRISMA Cluster of Excellence and Johannes-Gutenberg University Mainz (Germany)\\
        E-mail: \email{doria@uni-mainz.de}}

\author{Patrick Achenbach\\
        PRISMA Cluster of Excellence and Johannes-Gutenberg University Mainz (Germany)\\
        Helmholtz Institute Mainz\\
        E-mail: \email{achenbach@uni-mainz.de}}

\author{Mirco Christmann\\
        PRISMA Cluster of Excellence and Johannes-Gutenberg University Mainz (Germany)\\
        E-mail: \email{mircochr@students.uni-mainz.de}}

\author{Achim Denig\\
        PRISMA Cluster of Excellence and Johannes-Gutenberg University Mainz (Germany)\\
        Helmholtz Institute Mainz\\
        E-mail: \email{denig@uni-mainz.de}}

\author{Harald Merkel\\
        PRISMA Cluster of Excellence and Johannes-Gutenberg University Mainz (Germany)\\
        E-mail: \email{merkel@uni-mainz.de}}

\abstract{Dark Matter is being searched with a variety of methods, each of which tackles this challenge
  focusing on different kinds of particles, masses and couplings. Here we describe Dark Matter searches
  conducted with accelerators and fixed targets. In particular, we highlight the contribution of the
  experiments being built at the new Mainz Energy-recovery Superconducting Accelerator (MESA) facility.
  MESA will provide intense electron beams for hadron and nuclear physics, as well as for Dark Matter searches
  with competitive sensitivities.}

\FullConference{ALPS 2019 An Alpine LHC Physics Summit \\
		April 22 - 27, 2019\\
		Obergurgl, Austria}

\begin{document}

\section{Introduction}
A variety of astrophysical and cosmological observations point to the existence of Dark Matter (DM).
In the last decades, an intensive and complementary effort was made searching for DM with particle colliders,
fixed-target experiments, indirect detection techniques, and underground direct detection experiments with negative results. Particle DM
remains still elusive, and more sensitive experiments are required.
A large class of models describes DM as a relic from the early Universe where DM was in thermodynamic
equilibrium with Standard Model (SM) particles.
DM abundance was set when its annihilation rate in SM particles became smaller than the expansion rate of the universe
set by the Hubble constant, a process known as {\em freeze-out}. Although this mechanism is very compelling, it allows for
a very broad range of DM masses (keV/c$^2$-TeV/c$^2$) and interaction cross-sections.
For long time, the target of many experiments was a DM candidate
called WIMP (weakly-interacting massive particle), motivated by models of physics beyond the SM
(most notably supersymmetric models).
This range can be effectively tackled by high-energy colliders and direct detection experiments.
In the {\em light dark matter} $<$GeV/c$^2$ range (LDM) these methods become less effective, and different techniques are needed.
%with the lowest thresholds for DM-nucleon scattering obtained with cryogenic silicon and germanium detectors \cite{scdms}.
%Experiments at high-energy colliders are not optimized for the low mass range: a typical DM
%signature would be the detection of missing mass, which is too small to be detected
%given the backgrounds present.
%Direct detection experiments are mostly based on the nuclear recoil signal from galactic DM scattering,
%and for low masses it becomes too small to be detected, although
%more recently, notable progress has also been made on DM-electron scattering \cite{xenon,ecdms}.
LDM production at accelerators can in principle overcome the limitations of direct detection experiments
(the detection energy threshold) by producing DM particles with enough momentum to be detected, even if
they are relatively light.
High-energy colliders are not optimized for the low-mass range, and lower energy accelerators coupled
with fixed targets can have an advantage in LDM searches.

\section{A Theoretical Model for Light Dark Matter}
In the $<$GeV/c$^2$ range, for retaining a thermal origin for DM, we have to postulate the existence of additional interactions.
For small masses, considering only electroweak-scale cross-sections the annihilation rate would not
be sufficient, leading to DM overproduction.
A relevant class of LDM models is based on the idea that DM particles belong to a {\em dark sector}
interacting with the SM via one (or more) mediator particle(s).
Dark sector models can be classified by the type of the mediator particle
(the {\em portals}) and the type of DM particle.
In general, the dark sector might contain more mediators and particles.
Here we focus on a simple model which still captures the essence of dark sector physics.
The model is comprised by a massive vector mediator particle (a "dark photon")
which can be thought as a gauge boson resulting from a spontaneously broken $U(1)_D$
symmetry. The model's lagrangian is
\begin{equation}
  \begin{aligned}
  \mathcal{L} &\supseteq -\frac{1}{4}F'_{\mu\nu}F'^{\mu\nu}+\frac{\epsilon_Y}{2}F'_{\mu\nu}B^{\mu\nu}\\
  &+\frac{m^2_{A'}}{2}A'_{\mu}A'^{\mu} + g_DA'_{\mu}J_{\chi}^{\mu}+g_YB_{\mu}J^{\mu}_Y \quad,
  \end{aligned}
\end{equation}
where $F'_{\mu\nu} = \partial_{\mu}A'_{\nu}- \partial_{\nu}A'_{\mu}$ and
$B_{\mu\nu} = \partial_{\mu}B_{\nu}- \partial_{\nu}B_{\mu}$ are
the dark photon and the hypercharge fields, $g_D$ is the dark gauge coupling,
and $J_{\chi}^{\mu}$ and $J^{\mu}_Y$ the DM and hypercharge currents, respectively.
After electroweak symmetry breaking, the dark photon mixes with the SM photon and the Z boson
\begin{equation}
  \frac{\epsilon_Y}{2}F'_{\mu\nu}B^{\mu\nu} \rightarrow \frac{\epsilon}{2}F'_{\mu\nu}F^{\mu\nu} +
  \frac{\epsilon_Z}{2}F'_{\mu\nu}Z^{\mu\nu} \quad,
\end{equation}
where $\epsilon=\epsilon_Y / \cos\theta_W$, $\epsilon_Z=\epsilon_Y/\sin\theta_W$, and
$\theta_W$ is the weak mixing angle. After diagonalization, the coupling of the dark
photon with DM and the SM photon is
\begin{equation}
  g_DA'_{\mu}J_{\chi}^{\mu}+g_YB_{\mu}J^{\mu}_Y \rightarrow A'_{\mu}(g_DJ_{\chi}^{\mu}+\epsilon e J_{EM}^{\mu}) \quad,
\end{equation}
where $J_{EM}^{\mu}$ is the SM electromagnetic current. The form of the DM current $J_{\chi}^{\mu}$ depends
on the exact nature of the DM particle. The coupling of the dark photon to SM particles happens
via the "millicharge" $\epsilon e$, while the dark fine structure constant $\alpha_D=\sqrt{4\pi g_D}$
describes the coupling with DM.\\
It is useful to define the dimensionless combination of the model parameters
\begin{equation}
  y = \epsilon^2\alpha_D\left( \frac{m_{\chi}}{m_{A'}} \right)^4 \quad,
\end{equation}
which is proportional to the thermally averaged DM annihilation rate. 

\section{High Intensity Accelerators}
The search for very rare events like DM or neutrino interactions needs intense beams.
A typical figure of merit for an accelerator is its power $P(W) = I(A)\cdot E(eV)\cdot w$, where $E$ is the beam energy,
$I$ is the instantaneous current and $w$ the duty cycle.
The beam power is important because on the one hand it represents the amount of power that
the RF system has to deliver to the beam, and on the other hand it corresponds to its efficiency
for the production of secondary particles ({\em e.g.}, neutrons, pions, kaons).
For the production of high-power beams, linear accelerators have advantages as compared
with other types of accelerators: the repetition frequency of a linac is not limited by the rise time of
the magnets as is the case for synchrotrons.

In experiments based on the timing coincidence of two detectors, another key feature of modern
accelerators is the duty cycle $w$, which is the ratio between the length of the bunch of
particles and the distance between bunches. If $w=1$, the beam is ``continuous'' ($cw$: continuous wave)
and every RF bucket is filled with particles. Achieving $cw$ operation with normal conducting linacs requires
huge amounts of power. Modern accelerators achieve $cw$ beams with superconducting accelerating cavities
and/or recirculation, as in the case of CEBAF at JLab\footnote{Thomas Jefferson National Accelerator Facility,
  12000 Jefferson Avenue, Newport News, VA}
and MAMI\footnote{Mainzer Mikrotron, Inst. f\"ur Kernphysik Johannes Gutenberg-Universit\"at Mainz, J. Becher Weg 45 D 55128 Mainz} in Mainz.

\begin{figure}[t!]
  \begin{center}
    \includegraphics[width=\textwidth]{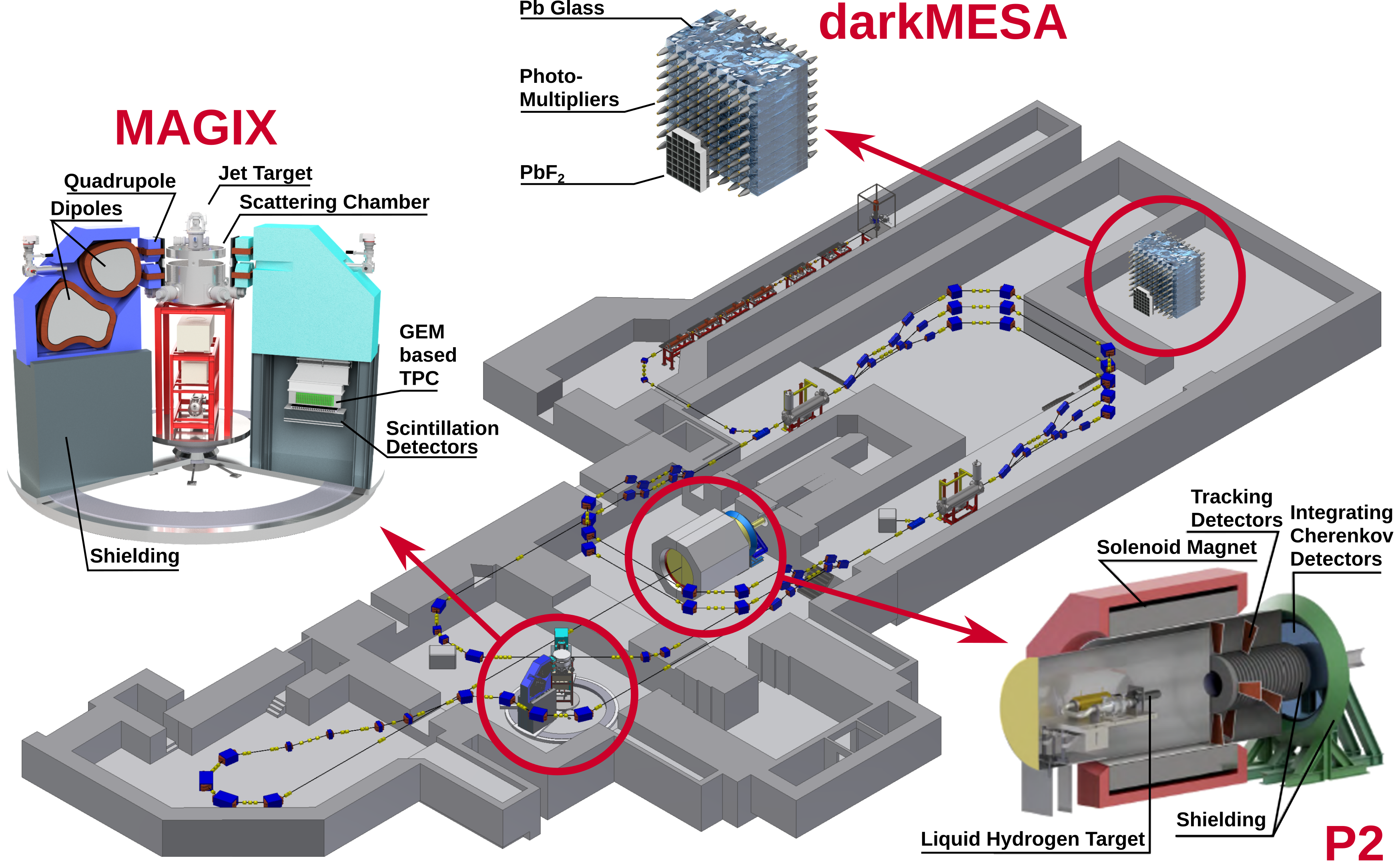}
  \end{center}
  \caption{The MESA (Mainz Energy-Recovery Superconducting Accelerator) complex with the three foreseen experiments: P2, MAGIX, and DarkMESA.}
  \label{fig:MESA}
\end{figure}

\section{The MESA accelerator}
The Institute for Nuclear Physics at the Mainz University is building a new
$cw$ multi-turn energy recovery linac for precision particle physics experiments
with a beam energy range of 100-200 MeV.
MESA will operate in two modes: energy recovery mode (ERM) and external beam mode (XBM).
In ERM, the accelerator will provide a beam current of up to 1~mA at 105 MeV for the MAGIX internal
target experiment with multi-turn energy recovery capability.
In XBM, a polarized beam of 150 $\mu A$ will be provided to the P2 experiment \cite{P2}.
The linac will provide an energy gain of 50 MeV/pass by using four ELBE-like 9-cell cavities \cite{cavities}
installed in two cryomodules.

\section{Decays of the Dark Photon}
If the dark photon can decay into SM particles ({\em e.g.} $e^+e^-$, $\mu^+\mu^-$), these are called {\em visible decays}.
In this case it is possible to detect the decay particles in coincidence and reconstruct the dark photon mass.
Fixed target experiments based on different techniques are operating or will be ready in the near future for visible decay searches.
At JLab, two experiments are operating. The first, HPS \cite{HPS}, leverages on advanced tracking
capabilities with silicon microstrip detectors for reconstructing the decay particles. The second, APEX \cite{APEX}, is based
on two high-resolution magnetic spectrometers. Magnetic spectrometers were used also by the A1 Collaboration at MAMI \cite{A1}.
If the dark photon is heavier than twice the DM particle, the decay $\gamma'\rightarrow \chi\chi$ is possible: this is
the case of {\em invisible decays}. Missing mass or missing momentum experiments can search for invisible decays, as well as
beam-dump experiments.\\
The projected LDMX experiment will have an extremely high sensitivity with a technique based on missing
momentum reconstruction \cite{LDMX}. An example of beam-dump experiment is BDX, which is planned at JLab \cite{BDX}. 
For a review of the different experimental approaches, refer for example to \cite{cvisions17}.

\section{The MAGIX Experiment}
MAGIX is a flexible experiment exploiting the unique combination of a supersonic gas-jet target and
the MESA $cw$ beam in energy recovery mode.
The experiment is based on two magnetic spectrometers which can be operated in coincidence
with momentum resolution $\delta p / p \sim 10^{-4}$ and low-material budget focal plane detectors.
MAGIX will allow precision measurements in a variety of fields
ranging from hadron physics to nuclear astrophysics and dark sector searches.
The dark photon can be produced through a mechanism similar to bremsstrahlung on a heavy
nuclear target Z via the reaction $e^-Z\rightarrow e^-Z\gamma^{\prime}$. If the dark photon decays
into SM particles, {\em e.g.} $\gamma^{\prime}\rightarrow e^+e^-$, the electron/positron final state 
can be detected in coincidence in the two spectrometers and a peak-search on the QED background can be
performed. If the dark photon decays invisibly ({\em e.g.} into light dark matter particles $\gamma^{\prime}\rightarrow \chi\bar{\chi}$),
this will require the measurement of the recoil target nucleus for fully reconstructing the kinematics.
A peak-search on the reconstructed missing mass $m_{\gamma^{\prime}}^2=(p_{beam}-p_{Z}-p_{e'})^2$ will be performed in this case
with the addition of a silicon detector for detecting the recoil proton or nucleus.
Fig.~\ref{fig:simulations} (left) shows the projected sensitivity of MAGIX to the dark photon visible decays.

\section{The DarkMESA beam-dump Experiment}
In a beam-dump experiment, the dark photon can be produced radiatively by an impinging electron beam on a
heavy nucleus Z through the bremsstrahlung-like process $eZ\rightarrow eZ\gamma'$ \cite{Bjorken:2009}.
We assume that the dark photon decays into pairs of DM particles ($\gamma' \rightarrow \chi\bar{\chi}$).
Depending on the model, $\chi$ and $\bar\chi$ can be a particle/antiparticle couple or two
different particles $\chi_1$ and $\chi_2$ (inelastic DM \cite{iDM}).\\
After production, DM particles can be detected with a shielded detector downstream of the
beam-dump via $e\chi \rightarrow e\chi$ and $p\chi \rightarrow p\chi$ scattering, where $p$
is a proton.\\
The dark photon production yield scales as $Y_{\gamma'} \sim \alpha^3\epsilon^2 / m^2_{A'}$
while the DM yield $Y_{\chi}$ in the detector is proportional to $\epsilon^2$, giving
a total number of detected DM particles scaling as $Y_{\gamma'}\cdot Y_{\chi} \sim \epsilon^4$. While the detection
yield does not have a favorable scaling, a beam-dump experiment has distinct advantages.
The large number of electrons on target (EOT) deliverable in a reasonable amount
of time by modern $cw$ electron accelerators can compensate for the small yield
and reach high sensitivity.
Another advantage is provided by the boost at which DM particles are produced, allowing
for an improved reach at low masses.
Moreover, such experiments are unique since they can probe at the same time both the dark
photon production/decay and the DM interaction with SM particles.
At MESA, a radiation-shielded area is available 23~m downstream of the beam-dump of the P2 experiment,
allowing for the installation of a detector for LDM searches (see Fig.~\ref{fig:MESA}).
The current plan for the construction of the experiment is divided into three phases.
Phase-1 will employ already available PbF$_2$ crystals for building a $(1\times 1\times 0.13)$~m$^3$ detector.
Phase-2 will add Pb-Glass crystals expanding the Phase-1 detector, while in Phase-3 the full available
volume will be exploited, reaching a total detector volume of  $(2.7\times 2.7\times 1.5)$~m$^3$.
The advantage of Cherenkov crystals is their speed and relatively low sensitivity to background neutrons.
Taking advantage of the $3\times 10^{22}$ electrons on target delivered to the P2 experiment
at 150~$\mu$A of beam current, a total charge of $\sim$5400~C will be deposited in the beam-dump and will
be available for DarkMESA.

\begin{figure}[!t]
  \centering
\begin{minipage}[t]{0.49\textwidth}
\includegraphics[width=1.0\textwidth]{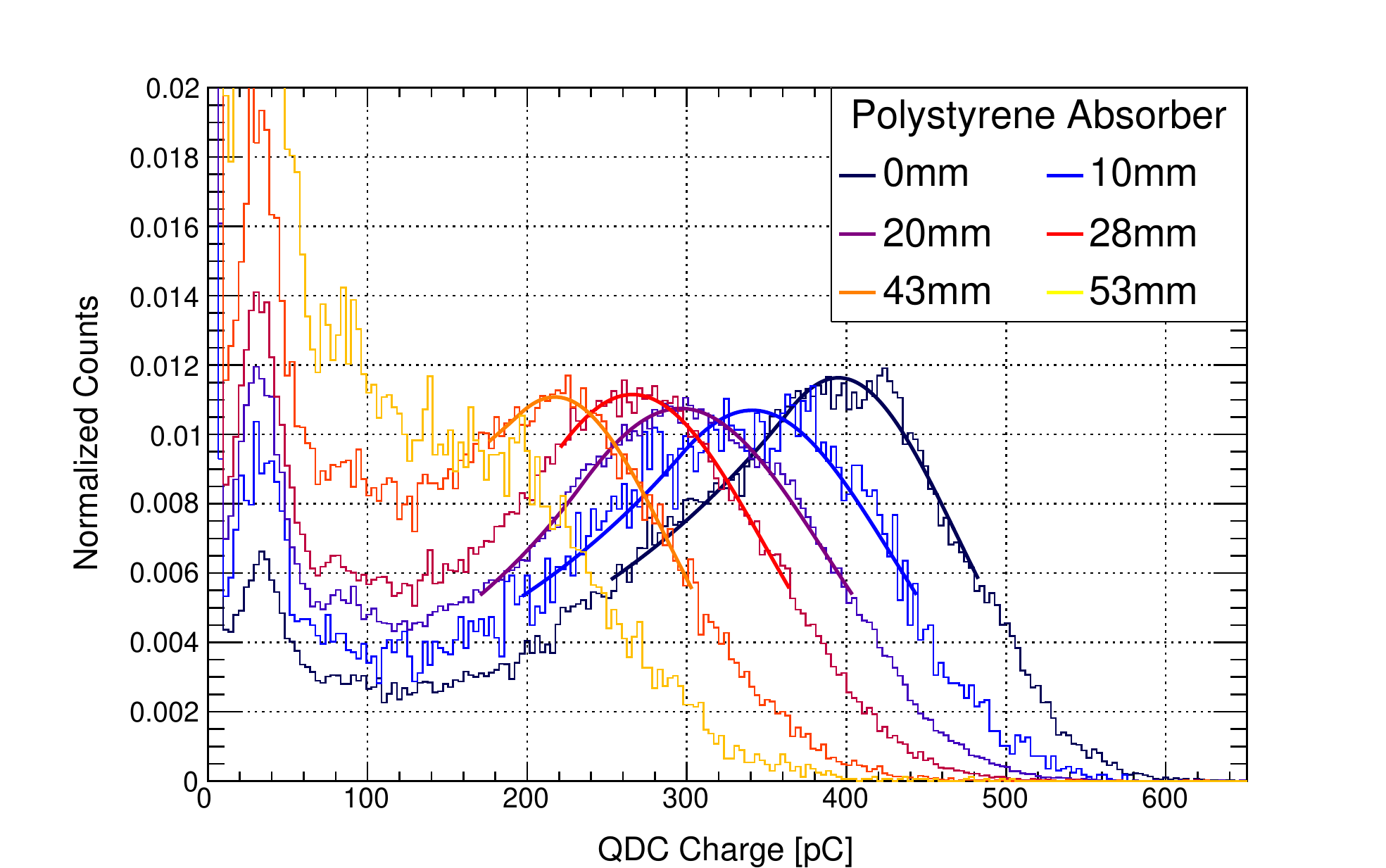}
\end{minipage}
\begin{minipage}[t]{0.49\textwidth}
  \includegraphics[width=0.872\textwidth]{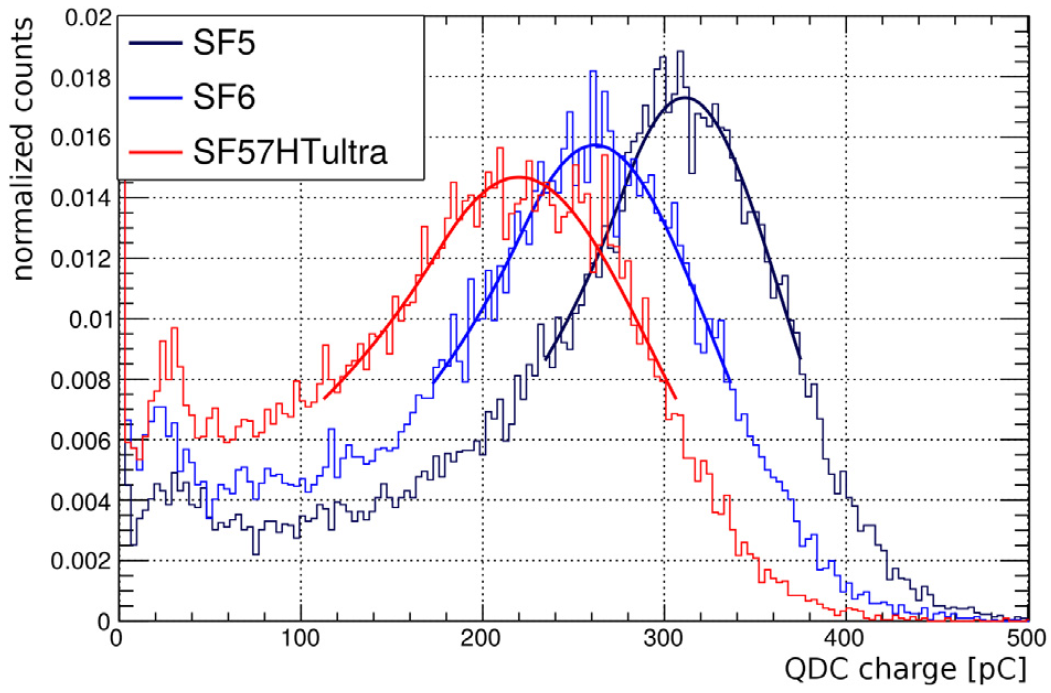}
\end{minipage}
\caption{{\bf (Left)} QDC spectra collected at the MAMI 14~MeV beamline with PbF$_2$ crystals with different plastic absorbers.
  The energy loss of the electron beam in the absorbers is 2~MeV/cm without significant energy straggling.
  {\bf (Right)} QDC spectra corresponding to the tested lead-glasses at 14~MeV beam energy.}
\label{fig:detector1}
\end{figure}

\newpage
\section{DarkMESA Detector Development}
\begin{figure}[!t]
  \centering
  \includegraphics[width=1.0\textwidth]{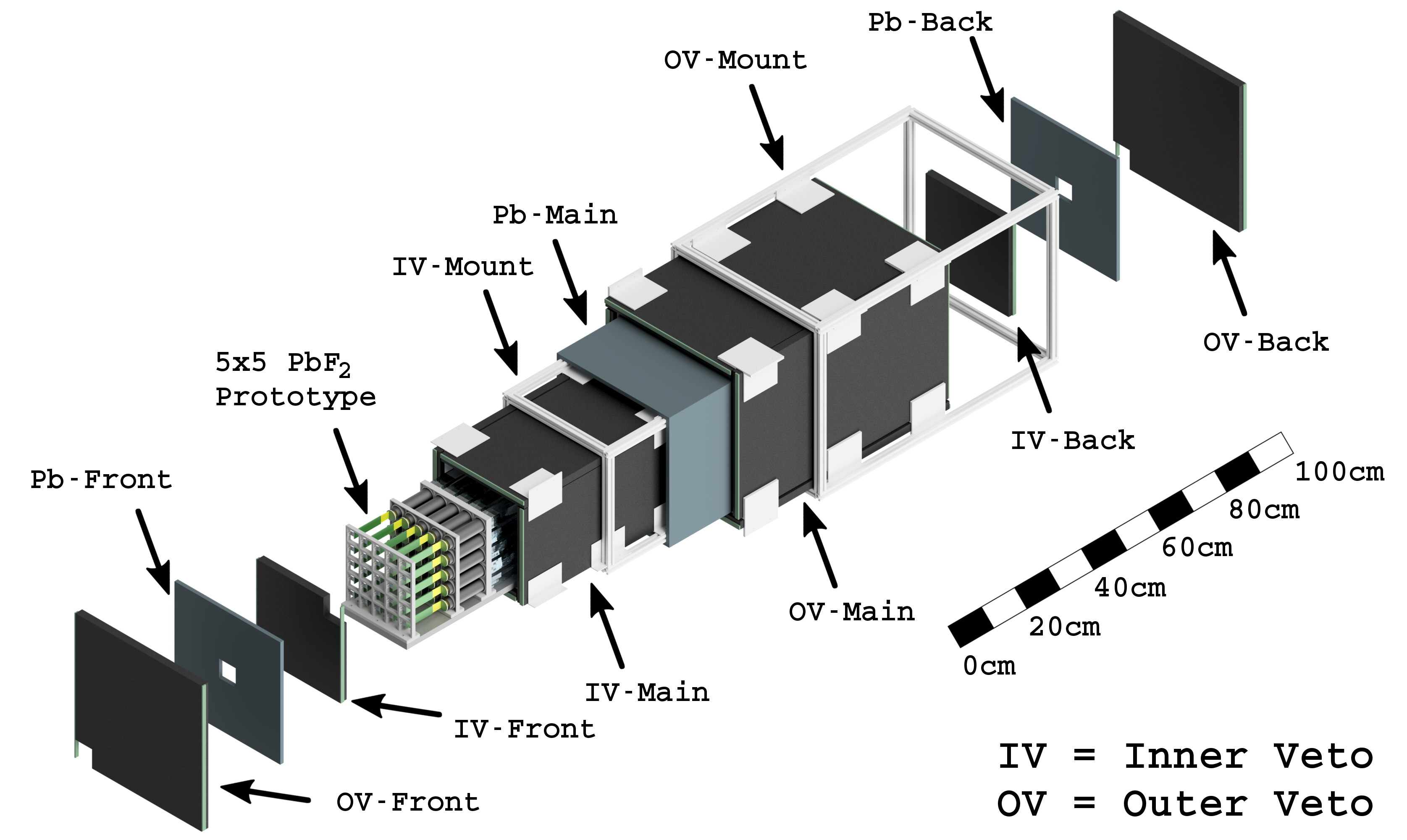}
  \caption{Open view of the DarkMESA prototype module: the PbF$_2$ crystals and the corresponding PMTs are
  installed in a 3D-printed structure which is enclosed in a cosmics veto system. The veto system is composed by
  two plastic scintillator layers separated by a lead layer. }
  \label{fig:detector2}
\end{figure}

Detector development for DarkMESA has already started with the characterization of the  PbF$_2$ and Pb-Glass crystals
and the construction of a $5\times 5$ crystals detector prototype equipped with a cosmics veto system (see Fig.~\ref{fig:detector2}).
The detector materials studied were PbF$_2$, and the Schott\footnote{SCHOTT AG, Hattenbergstrasse 10, 55122, Mainz (Germany)}
Pb glasses SF5, SF6, SF57HTultra.
Beam tests at MAMI were performed with a 14~MeV beam \cite{mirco}. Polystyrene absorbers (average energy loss $\sim$ 2 MeV/cm)
were used to characterize the detectors at lower energies.
A fiber detector was used as trigger and beam position monitor. Photonis XP2900 photomultipliers with 1 1/8''
diameter were used and spectra were recorded with a CAEN V965 QDC. X-Y scans on the front ($0^{\circ}$ with respect of the beam axis)
and side surface ($90^{\circ}$) were performed, and also the $45^{\circ}$ incident angle was examined.
Selected results from the beam test are showed in Fig.~\ref{fig:detector1}.

SF5 was found the best lead-glass tested, while PbF$_2$ produces more light than all the lead-glasses.
The experimental results are in good agreement with {\tt Geant4} optical simulations.
The detector efficiency for electrons was calculated with a simulation. For both SF5 and PbF$_2$ the detection
efficiency for 10 MeV electrons, assuming that at least 5 photoelectrons were detected, is above 90\%.
The response of the crystals to low-energy neutrons was also studied. Using a neutron source, the detection efficiency
was found to be O($10^{3}$) lower with respect to a reference plastic scintillator.

%\begin{figure}[!t]
%  \centering
%\begin{minipage}[t]{0.49\textwidth}
%\includegraphics[width=1.0\textwidth]{prototype.png}
%\end{minipage}
%\begin{minipage}[t]{0.49\textwidth}
%  \includegraphics[width=1.0\textwidth]{neutron.png}
%\end{minipage}
%\caption{{\bf (Left)} Open view of the DarkMESA prototype module: the PbF$_2$ crystals and the corresponding PMTs are
%  installed in a 3D-printed structure which is enclosed in a cosmics veto system. The veto system is composed by
%  two plastic scintillator layers separated by a lead layer. 
%  {\bf (Right)} Neutron detection efficiency with respect to a plastic scintillator for the tested Cherenkov crystals.}
%\label{fig:detector2}
%\end{figure}

\section{Simulation}
\begin{figure}[!t]
  \centering
\begin{minipage}[t]{0.48\textwidth}
\includegraphics[width=1.0\textwidth]{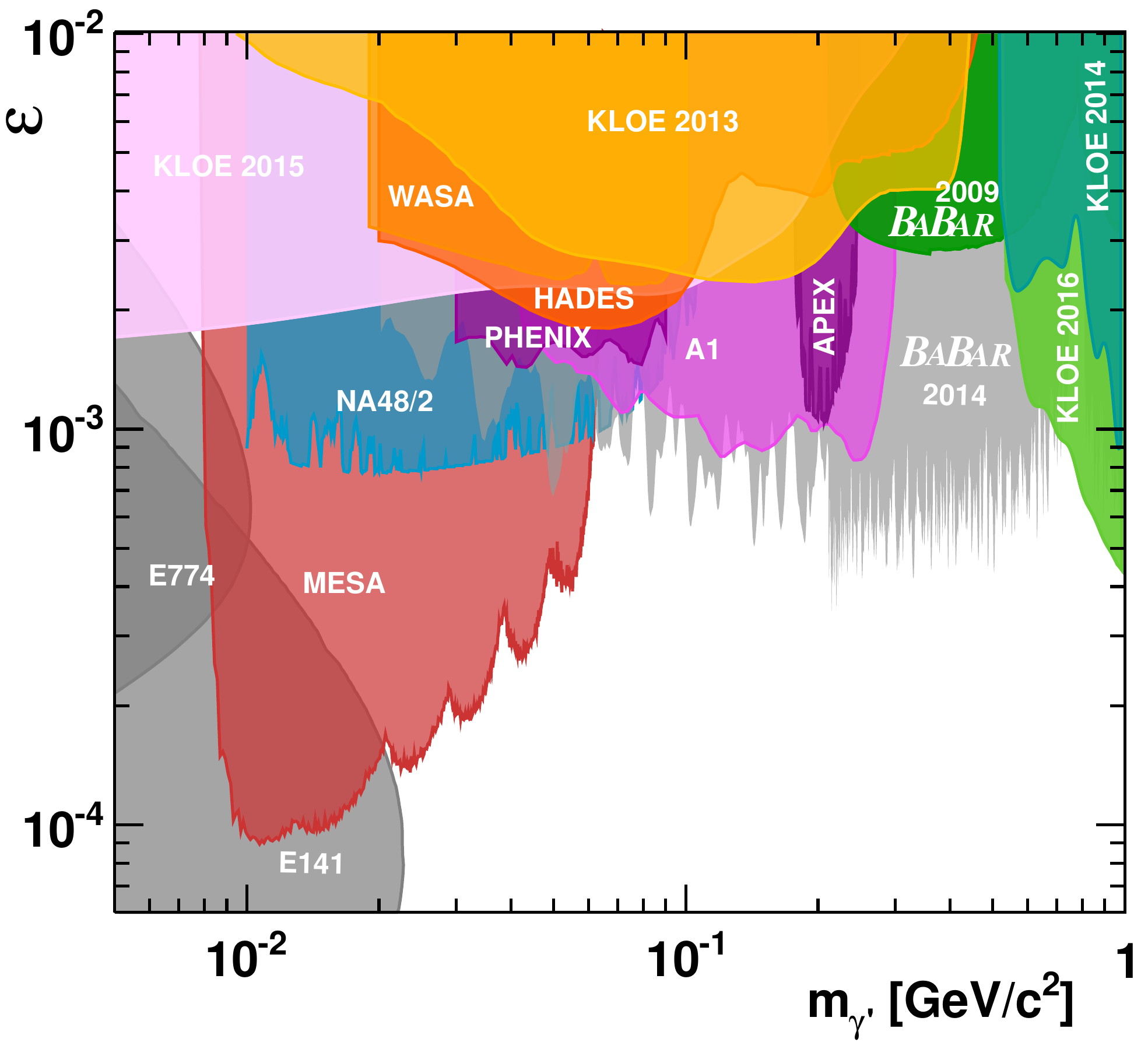}
\end{minipage}
\begin{minipage}[t]{0.50\textwidth}
  \includegraphics[width=1.0\textwidth]{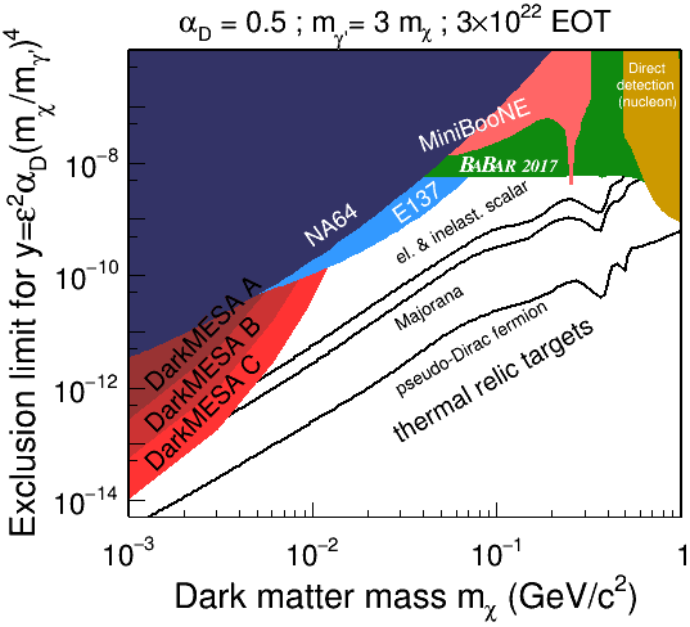}
\end{minipage}
\caption{{\bf (Left)} MAGIX projected exclusion limits for $\epsilon$ as function of the dark photon mass for the $\gamma'\rightarrow e^+e^-$ decay.
  {\bf (Right)} DarkMESA projected exclusion limits for the $y$ parameter as a function of the DM particle mass for the
  invisible dark photon decay $\gamma'\rightarrow \chi\bar{\chi}$.}
\label{fig:simulations}
\end{figure}

For assessing the sensitivity of a beam-dump experiment at MESA, a full simulation study was performed \cite{cipanp}.
The {\tt Geant4} \cite{GEANT} simulation implemented the geometry of the experimental halls, the
relevant details of the P2 experiment, the beam-dump, and the detector.
{\tt MadGraph4} \cite{MG4} was used for generating the dark photon bremsstrahlung process.
In the simulation, a mono-energetic 155~MeV pencil electron beam hits
the 60-cm long P2 liquid hydrogen target. The $\sim$0.6~T solenoidal magnetic field of P2 was simulated and its
effect resulted in focusing the particles emerging from the target on the beam-dump which is made of aluminum
and cooling water.\\
Fixing the parameters of the theoretical model ($m_{\gamma'}$, $m_{\chi}$, $\epsilon$, $\alpha_D$),
the dark photon production cross section and the DM final state four-vectors were calculated with
{\tt MadGraph4}. The showering effects in the beam-dump were fully taken into account.
The final state four-vectors for the $\chi /\bar{\chi}$ particles were re-introduced in the {\tt Geant4} simulation
where they were tracked through the various materials up to the detector location.
The $\chi /\bar{\chi}$ interaction with electrons or protons in the detectors was calculated
with a custom code embedded into {\tt Geant4} implementing the $e\chi$ and $p\chi$ scattering cross-sections
at first order dark photon exchange.
The total number of detected DM particles was calculated as
\begin{equation}
  \begin{aligned}
    N_{\chi\bar{\chi}}  = EOT \times N_{D} \times N_{DET} \times N_{BD}\\
    \times X_0 \times \frac{\sigma_{MG}}{N_{SIM}} \times \sum_{i=0}^{i=N_D}L_i\sigma_i \quad,
  \end{aligned}
  \label{Nchifinal}
\end{equation}
where EOT is the number of electrons on target, $N_{D}$ is the number of $\chi /\bar{\chi}$ within the detector acceptance, $X_0$
the beam-dump radiation length, $L_i$ the track length in the detector of the $i-th$ DM particle track, $\sigma_i$
the $e\chi\rightarrow e\chi$ or $p\chi\rightarrow p\chi$ cross section of the $i-th$ DM particle track, $\sigma_{MG}$ the
$eA\rightarrow eA\gamma'\rightarrow \chi\bar{\chi}$ cross-section calculated with {\tt MadGraph4}, and $N_{SIM}$ the total
number of simulated events.
For a detector with a combination of materials with average atomic number Z, mass number A, and density $\rho_D$,
the total number of scattering centers (number of electrons or protons) is $N_{DET}=Z\rho_DN_A/A$,
where $N_A$ is the Avogadro number. With the same notation, the number of nuclei in the beam dump is $N_{BD}=\rho_{BD}N_A/A$.
The results of the simulation for the two phases are summarized in Fig.~\ref{fig:simulations} (right) together with
the existing limits \cite{na64-1,na64-2,babar,minib,cresst}.
The 3-$\sigma$ exclusion limits on the thermal target variable $y$ are calculated
as a function of the dark matter mass $m_{\chi}$ conservatively assuming $m_{\gamma^{\prime}}=3m_{\chi}$,
$\alpha_D=0.5$ and a detector threshold $E_{min}$=14~MeV.  The experiment requires
an efficient cosmics veto system and the detector threshold should be high enough in order
to exclude natural radioactivity backgrounds. The experiment will work below the pion production
threshold, thus neutrino or muon backgrounds from the beam-dump will be absent.

\section{Summary}
The new MESA accelerator at the Institute for Nuclear Physics of the Johannes Gutenberg University in Mainz
will allow new exciting opportunities in precision tests of the Standard Model, nuclear physics,
as well as in new physics searches connected to the long-standing dark matter problem.
The MAGIX two-spectrometer setup, exploiting the high luminosity provided by MESA in combination
with a gas-jet target, will be able to search for visible and invisible decays of the dark photon in a
new mass range.
The installation of a beam-dump experiment represents an unique opportunity to expand the MESA research program
with a competitive experiment working parasitically to the other ones, taking
advantage of the world-class EOT delivered by the new accelerator.
Moreover, beam-dump experiments have the advantage of being able to investigate at the
same time the production of the dark photon, its decay, and the DM interaction.

\end{document}